\documentclass[10pt, conference]{IEEEtran}
\IEEEoverridecommandlockouts

\usepackage{cite}
\usepackage{amsmath}
\usepackage{amssymb}
\usepackage{amsfonts}
\usepackage{amsthm}
\usepackage{comment}
\usepackage{algorithmic}
\usepackage{graphicx}
\usepackage{textcomp}
\usepackage{xcolor}
\usepackage{dsfont}
\usepackage{hyperref}
\usepackage{quantikz}
\usepackage{subcaption}
\usepackage{placeins}
\usepackage{float}
\usepackage{multirow}
\usepackage{booktabs}
\usepackage{balance}

\theoremstyle{definition}
\newtheorem{example}{Example}[section]

\begin{document}

\title{Higher-Order Portfolio Optimization with Quantum Approximate Optimization Algorithm}

\author{\IEEEauthorblockN{Valter Uotila}
\IEEEauthorblockA{\textit{Aalto University} \&  \textit{University of Helsinki}\\
valter.uotila@aalto.fi}
\and
\IEEEauthorblockN{Julia Ripatti}
\IEEEauthorblockA{\textit{Aalto University} \\
julia.1.ripatti@aalto.fi}
\and
\IEEEauthorblockN{Bo Zhao}
\IEEEauthorblockA{\textit{Aalto University} \\
bo.zhao@aalto.fi}
}

\maketitle

\begin{abstract}
Portfolio optimization is one of the most studied optimization problems at the intersection of quantum computing and finance. In this work, we develop the first quantum formulation for a portfolio optimization problem with higher-order moments, skewness and kurtosis. Including higher-order moments leads to more detailed modeling of portfolio return distributions. Portfolio optimization with higher-order moments has been studied in classical portfolio optimization approaches but with limited exploration within quantum formulations. In the context of quantum optimization, higher-order moments generate higher-order terms in the cost Hamiltonian. Thus, instead of obtaining a quadratic unconstrained binary optimization problem, we obtain a higher-order unconstrained binary optimization (HUBO) problem, which has a natural formulation as a parametrized circuit. Additionally, we employ realistic integer variable encoding and a capital-based budget constraint. We consider the classical continuous variable solution with integer programming-based discretization to be the computationally efficient classical baseline for the problem. Our extensive experimental evaluation of 100 portfolio optimization problems shows that the solutions to the HUBO formulation often correspond to better portfolio allocations than the classical baseline. This is a promising result for those who want to perform computationally challenging portfolio optimization on quantum hardware, as portfolio optimization with higher moments is classically complex. Moreover, the experimental evaluation studies QAOA's performance with higher-order terms in this practically relevant problem.
\end{abstract}


\section{Introduction}

A key goal of quantum algorithm development is to identify and construct well-motivated problems that can be solved faster or more efficiently on quantum computers than on classical hardware in real-life applications. At the intersection of quantum computing and finance, portfolio optimization is one of the most studied and promising problems in this regard \cite{Buonaiuto_Gargiulo_De_Pietro_Esposito_Pota_2023, Brandhofer_Braun_Dehn_Hellstern_Huls_Ji_Polian_Bhatia_Wellens_2022, Herman_Googin_Liu_Galda_Safro_Sun_Pistoia_Alexeev_2022, Orus_Mugel_Lizaso_2019, GALLUCCIO1998449, Venturelli_Kondratyev_2019, Rosenberg_Haghnegahdar_Goddard_Carr_Wu_Prado_2016, Rebentrost_Lloyd_2018, Phillipson_Bhatia_2020, Kerenidis_Prakash_Szilagyi_2019, Hegade_Chandarana_Paul_Chen_Albarran_Arriagada_Solano_2022, Hodson_Ruck_Ong_Garvin_Dulman_2019, Marzec_2014, Mugel_Kuchkovsky_Sanchez_Fernandez_Lorenzo_Luis_Hita_Lizaso_Orus_2022, Grant_Humble_Stump_2021, Slate_Matwiejew_Marsh_Wang_2021, Baker_Radha_2022}. Some of the reasons for its popularity are its central role in finance and the fact that it has a straightforward quantum computational formulation as a quadratic unconstrained binary optimization (QUBO) problem. This article presents a more advanced higher-order portfolio optimization problem for quantum computers and benchmarks its performance with the standard \emph{quantum approximate optimization algorithm} (QAOA) \cite{farhi2014quantum}. The problem formulation includes encoding of higher-order moments: skewness and kurtosis. Additionally, we consider realistic integer variables and a capital-based budget constraint.

Our problem formulation can be divided into two parts: the higher-order unconstrained binary optimization formulation and the classical continuous variable formulation. A solution to both problems is an allocation, which describes how many assets an investor should buy, staying close to the budget. Comparing the expected returns of the allocations and the leftover budgets provides precise metrics for evaluating these two methods. Higher-order unconstrained binary optimization problems are eigenvalue problems that can be solved with QAOA. However, QAOA has not been extensively experimentally evaluated in these types of real-world higher-order optimization use cases. The classical continuous variable solution is a standard classical algorithm to solve portfolio optimization problems \cite{Martin2021}. It is efficient because it utilizes continuous variables, which are generally easier to optimize than integer variables \cite{Thomaser_De_Nobel_Vermetten_Ye_Back_Kononova_2023, 10.1007/s10107-022-01862-z}.


The contributions of this work are as follows.
\begin{itemize}
    \item We develop the first quantum portfolio optimization problem that includes higher moments (skewness and kurtosis), making the optimization problem more realistic.
    \item Our evaluation shows that the solution quality of the higher-order binary optimization formulation is often better than the solution quality of the classical continuous variable formulation with integer programming-based discretization. This finding motivates further research of higher-order binary optimization with quantum methods.
    \item The experimental evaluation benchmarks QAOA on 100 instances of higher-order binary portfolio optimization problems and identifies several challenges associated with its application.
\end{itemize}

\paragraph{Related work on quantum and finance} Finance is one of the key areas where quantum computing is believed to produce value even in the near future \cite{Orus_Mugel_Lizaso_2019, Herman_Googin_Liu_Galda_Safro_Sun_Pistoia_Alexeev_2022, Egger_Gambella_Marecek_McFaddin_Mevissen_Raymond_Simonetto_Woerner_Yndurain_2020}. So far, portfolio optimization has been the most studied problem in finance with quantum computing \cite{Buonaiuto_Gargiulo_De_Pietro_Esposito_Pota_2023, Brandhofer_Braun_Dehn_Hellstern_Huls_Ji_Polian_Bhatia_Wellens_2022, Cohen_Khan_Alexander_2020, Herman_Googin_Liu_Galda_Safro_Sun_Pistoia_Alexeev_2022, Orus_Mugel_Lizaso_2019, Venturelli_Kondratyev_2019, Rosenberg_Haghnegahdar_Goddard_Carr_Wu_Prado_2016, Rebentrost_Lloyd_2018, Phillipson_Bhatia_2020, Kerenidis_Prakash_Szilagyi_2019} and its connection to spin glass energy minimization problems was suggested as early as 1998 \cite{GALLUCCIO1998449}. Interestingly, higher-order polynomials or higher-order problems have not been discussed in the literature, except in \cite{Rosenberg_Haghnegahdar_Goddard_Carr_Wu_Prado_2016}, where they are briefly mentioned in the context of developing more suitable penalty functions for the budget constraint.

Table~\ref{tab:quantum_optimization_summary} reveals that no previous work has benchmarked portfolio optimization with QAOA using the realistic capital-based budget constraint and integer variables as we do in this work. None of the earlier works have employed higher-order moments. This is also the first time classical continuous variable portfolio optimization with an integer programming-based discretization has been compared to quantum formulations.

\begin{table}[t]
\caption{Quantum portfolio optimization approaches. If some works have not evaluated problems or used only a small number of them, this indicates that the research focus has been theoretical.}
\label{tab:quantum_optimization_summary}
\centering
\resizebox{\columnwidth}{!}{%
\begin{tabular}{@{}lcccc@{}}
\toprule
Paper & Integer vars & Capital budget & \#Problems evaluated & Method \\
\midrule
\cite{Rosenberg_Haghnegahdar_Goddard_Carr_Wu_Prado_2016} & Yes & No & - & Annealing \\
\cite{Hegade_Chandarana_Paul_Chen_Albarran_Arriagada_Solano_2022} & Yes & Yes & 1000 & Counterdiabatic \\
\cite{Grant_Humble_Stump_2021} & Yes & Yes & 4 & Annealing \\
\cite{Brandhofer_Braun_Dehn_Hellstern_Huls_Ji_Polian_Bhatia_Wellens_2022} & No & No & 20 & QAOA \\
\cite{Baker_Radha_2022} & No & No & 209 & QAOA \\
\cite{GALLUCCIO1998449} & Yes & Yes & 3 & Theory \\
\cite{Hodson_Ruck_Ong_Garvin_Dulman_2019} & No & No & 1 & QAOA \\
\cite{Venturelli_Kondratyev_2019} & No & No & 210 & Annealing \\
\cite{Mugel_Kuchkovsky_Sanchez_Fernandez_Lorenzo_Luis_Hita_Lizaso_Orus_2022} & Yes & No & 6 & Various \\
\cite{Buonaiuto_Gargiulo_De_Pietro_Esposito_Pota_2023} & Yes & Yes & 3 & VQE \\
\cite{Slate_Matwiejew_Marsh_Wang_2021} & No & No & 2 & Walks \\
\cite{Rebentrost_Lloyd_2018} & Yes & Yes & - & HHL \\
\cite{yalovetzky2024solving} & Yes & Yes & 6 & HHL$^{++}$ \\
\cite{qiskit_finance_portfolio_optimization} & No & No & - & QAOA \\
\cite{Phillipson_Bhatia_2020} & No & No & 20 & Annealing \\
\bottomrule
\end{tabular}
}
\end{table}

\paragraph{Related work on higher-order binary optimization} In this work, we apply a higher-order unconstrained binary optimization (HUBO) model \cite{boros_2002}. The HUBO problems resemble QUBO problems except that higher-degree interactions between the binary variables are allowed. Formally, given a binary variable vector $x = (x_1, \ldots, x_n)$ of length $n$ and the index set $[n] = \left\{1, \ldots, n \right\}$, the objective function is
\begin{equation}\label{eq:hubo_formal}
    f(x) = \sum_{S \subset V}\alpha_{S}\prod_{i \in S}x_{i},
\end{equation}
where $\alpha_{S} \in \mathds{R}$ and $S$ runs over all the subsets of $[n]$.

Without relying on quantum computing, HUBO problems can be addressed by using various classical methods, which often employ approximation methods, especially for larger problem sizes, due to the exponential nature of the problem. Integer programming methods can be utilized to approximate solutions to HUBO problems, but are limited by exponential complexity and do not guarantee exact feasibility \cite{munoz2017integer}. Different learning-based methods, such as graph neural networks, can approximate HUBO solutions \cite{schuetz2022combinatorial}, but they might not generalize well to unseen problems and can require significant training data. Previously, simulated annealing has been used to solve HUBO problems~\cite{wang2025speedup}. While these classical approaches remain competitive in solving small- to medium-scale problem sizes, they struggle with scalability, which could be potentially addressed with quantum native approaches. 

Solving HUBO problems on quantum computers is a small but growing area of interest in quantum optimization. Considering the previous work, \cite{campbell2022qaoa} studied solving a higher-order graph coloring problem with QAOA. Bias-field digitized counterdiabatic quantum optimization \cite{Romero_Visuri_Cadavid_Solano_Hegade_2024, Cadavid_Dalal_Simen_Solano_Hegade_2024} has been developed as a method to improve solving higher-order problems and there are methods to optimize variational circuits for
higher-order binary problems \cite{Verchere_Elloumi_Simonetto_2023}. Higher-order problems have also been used as a benchmark between QAOA and quantum annealing \cite{Pelofske_Bartschi_Eidenbenz_2023, Sachdeva_Hartnett_Maity_Marsh_Wang_Winick_Dougherty_Canuto_Chong_Hush_etal_2024, Pelofske_Bartschi_Eidenbenz_2024, Gilbert_Rodriguez_Louise_Sirdey_2023} and they have been used to understand the scaling of QAOA and its parameter concentration \cite{Pelofske_Bartschi_Cincio_Golden_Eidenbenz_2024}. Some initial applications of HUBO problems have been join order selection in relational database optimization \cite{uotila2025leftdeepjoinorderselection}, formalizing search for practical matrix multiplication algorithms \cite{10821361}, and optimizing railway rescheduling \cite{Domino_Kundu_Salehi_Krawiec_2022}. 

This paper begins by reviewing the classical mean-variance portfolio optimization problem and its extension to higher moments. We briefly discuss discretization via integer programming before formulating the quantum optimization approach, focusing on binary encoding and capital-based budgets. We then present the experimental setup, results, and visualizations, followed by conclusions and future directions. The implementation is available on GitHub \cite{github_repo}.
\section{Background}

We use consistent notation for variables across the paper: $w_i \in [0, 1]$ are the continuous variables in the unit interval, $z_i \in \mathds{Z}_{\geq 0}$ are non-negative integer variables, $x_i \in \left\{ 0, 1 \right\}$ are binary variables and $s_i \in \left\{ -1, 1 \right\}$ are spin variables.

\subsection{Markowitz mean-variance portfolio optimization}

In the original Markowitz mean-variance portfolio optimization problem \cite{Markowitz_1952}, we are given a set of assets and a budget. The goal is to invest the optimal amount in these assets to maximize the expected return or minimize the financial risk. The key information in portfolio optimization is the price data of the assets from a selected time period. Assume that we have $n$ assets indexed with $i \in [n] = \left\{1, \ldots, n \right\}$ and let $p^{t}_i$ denote the price of the asset $i$ at time $t$. By using this price data, the return $r_i^{t+1}$ is calculated for each asset between time $t$ and $t+1$ as 
\begin{equation}\label{eq:returns}
    r_i^{t+1}= \frac{p^{t+1}_i - p_i^t}{p_i^t}. 
\end{equation}


In real-life portfolio optimization, we are interested in the expected return values of assets over a selected time period, smoothing daily variation. Multiple methods exist to estimate the portfolio moments based on the returns in Eq.~\eqref{eq:returns}, and making accurate estimations is an active area of research. Some standard ways have been implemented in \cite{Martin2021}, and we rely on these regarding the first two moments. For each asset $i$, the mean returns can be computed either as an arithmetic mean
\begin{equation}\label{eq:expected_returns_arithmetic}
    \mu_{A, f}^{i} = \frac{f}{m}\sum_{t=1}^m r_i^t. 
\end{equation}
or as a geometric mean \cite{Martin2021}
\begin{equation}\label{eq:expected_returns_geometric}
    \mu_{G, f}^{i} =  \left(\prod_{t = 1}^{m}(1 + r_{i}^{t})\right)^{\frac{f}{m}} - 1,
\end{equation}
where $f$ is a so-called frequency. Usually, in finance, the mean return values are annualized. Therefore, we set the frequency as the number of trading days in a year, $f = 252$. In this work, we choose the geometric mean for calculating the mean return values since it is the default in \cite{Martin2021}.

The mean returns $\mu = (\mu_i)_{i=1}^{n}$ form the so-called first moment. The second moment is the covariance. The covariance between assets $i$ and $j$ can be defined as 
\begin{align}\label{eq:covariance}
    c_{ij} = \frac{f}{m - 1} \sum_{t=1}^m (r^{t}_{i} - \mu_{A, f=1}^{i})(r^{t}_{j}-\mu_{A, f=1}^{j}),
\end{align}
where we have taken into account annualization by scaling the regular covariance with a frequency of $f = 252$. The arithmetic mean $\mu_{A, f=1}^{i}$ is the mean value given in Eq.~\ref{eq:expected_returns_arithmetic} using frequency $f = 1$. The covariance matrix is $c = (c_{ij})_{i,j=1}^{m}$. We will define the third and fourth moments later. Before that, we briefly explain how the standard mean-variance portfolio optimization is formulated.

Let $w = (w_i)_{i \in [n]}$ be the vector of continuous variables. We assume that $w \geq 0$, which means that short selling is not allowed, which is a common assumption \cite{Brandhofer_Braun_Dehn_Hellstern_Huls_Ji_Polian_Bhatia_Wellens_2022}. The very first mean-variance portfolio optimization formulation aims to balance expected returns and risks so that either the returns are at least at a fixed level or the risks are at most at a fixed level. In the first case, the mean target return is fixed to $\mu_{\mathrm{fix}} \in \mathds{R}$, and we solve the following quadratic programming problem with linear constraints:
\begin{equation}\label{eq:program1}
\begin{array}{lll}
\mathrm{minimize}  & w^{\top} c w & \\
\mathrm{subject \ to} & w^{\top} \mu \geq \mu_{\mathrm{fix}}, & \\
                 & 1^{\top}w = 1. &
\end{array}
\end{equation}
The constraint $1^{\top}w = 1$ is the shorthand notation for $\sum_{i}w_i = 1$, which means that $100\%$ of the budget is invested. This program has a quantum formulation presented in \cite{Rebentrost_Lloyd_2018}. In other words, the program aims to minimize risk, as modeled using covariances, while maintaining a fixed level of returns. Alternatively, we can formulate an optimization program that maximizes the mean returns while keeping the risk at an acceptable level: 
\begin{equation}\label{eq:program2}
\begin{array}{lll}
\mathrm{maximize}  & w^{\top} \mu & \\
\mathrm{subject \ to} & w^{\top} c w \leq q_{\mathrm{fix}} , & \\
                 & 1^{\top}w = 1. &
\end{array}
\end{equation}
Here, the value $q_{\mathrm{fix}}  \in \mathds{R}$ is the value that controls the investor's risk level. Next, we further develop these formulations and connect them to quantum computing.

\subsection{Discretization}

The original portfolio optimization formulation is defined in terms of continuous weights $w_i \in [0, 1]$. The solution to the continuous variable problem indicates that an investor should invest $w_i \cdot 100\%$ of the budget in asset $i$. This creates a challenge, as we need to find a way to interpret the continuous weights as a discrete number of assets. This subsection reviews an integer programming-based method to discretize these continuous weights for a given budget \cite{Martin2021}.

In discretization, we need information about the current prices of the stocks. Let $p^{\tau} = (p^{\tau}_i)_{i=1}^{n}$ be the vector of closing prices from the last time moment $\tau$ to calculate the optimal allocation of stocks. In discretization, the goal is to convert the continuous variable weight vector $w \in [0,1]^{n}$ into a vector of integer variables $z \in \mathds{Z}_{\geq 0}^{n}$. Here, the integer vector $z$ represents the quantities of each stock to be purchased to achieve the optimal portfolio. Thus, the integer vector $z$ is the variable to be optimized in the discretization process. Let $C$ be the total capital and $C_{extra} = C - (p^{\tau})^{\top}z$ be the remaining unallocated capital. Then, the optimization problem is given by
\begin{equation}\label{eq:discretization}
\begin{array}{lll}
\mathrm{minimize}  & C_{extra} + |Cw - z^{\top}p^{\tau}| & \\
\mathrm{subject \ to} & C_{extra} + z^{\top}p^{\tau} = C, & \\
& z \in \mathds{Z}_{\geq 0}. & \\
\end{array}
\end{equation}
The solution to this optimization problem is an integer vector $z$ describing the discrete allocation of stocks. 


\subsection{Introducing discrete variables}

Asset prices and quantities are discrete in real life. Modeling them with continuous variables is computationally more efficient, but leads to certain problems and inaccuracies. Continuous variables may produce impractical solutions using real-life constraints, including minimum purchase sizes and indivisible assets. Continuous solutions might allocate small fractions of capital to many assets, which is infeasible in real markets due to transaction costs and indivisibility. The previous integer programming-based allocation does not necessarily fix these issues in practice because the discretization problem is separate from the optimization, which optimizes the weights.

In the following, we formulate the problem using integer variables, making the problem NP-hard. Instead of defining a weight vector $w = (w_i)_{i \in [n]}$, we define an integer variable vector $z = (z_i)_{i \in [n]}$ such that $z_i \in \mathds{Z}_{\geq 0}$. The interpretation of this vector is as follows: after optimizing the portfolio, we should have $z_i$ many assets of asset type $i$. This formulation eliminates the need for discretization, as we automatically obtain a discrete solution. The following subsections will express the classical portfolio optimization using integer variables. Later, we will explain how the integer variables can be mapped to binary variables, which are used in the quantum formulation.

\subsection{Unconstrained mean-variance portfolio optimization}

Compared to the previous linear and quadratic programs in Eq.~\eqref{eq:program1} and Eq.~\eqref{eq:program2}, we can also optimize the mean and variance simultaneously without limiting their values in the constraints. The weights should be optimized to maximize the mean return from the assets while minimizing the risk. The parameter $q_0 > 0$ controls the investor's willingness to take risks and the balance between risks and mean returns. In this case, the unconstrained portfolio optimization problem becomes
\begin{equation}\label{eq:unconstrained_problem}
\min_{z \in \mathds{Z}_{\geq 0}^{n}} q_0 z^{\top} c z - z^{\top} \mu.
\end{equation}
This unconstrained formulation assumes an unlimited, unrealistic budget, leading to an unbounded problem without a feasible solution. Therefore, we must include a budget constraint to obtain a viable solution.

\subsection{Budget constraint}

The budget constraint limits the total amount of capital that an investor can invest, and thus, it limits the number of assets that an investor can buy. This work assumes that the investor prefers to use the whole budget. 

Considering the continuous variable formulation, the problem with the budget constraint becomes
\begin{equation*}
\begin{array}{lll}
\mathrm{minimize}  & q_0 w^{\top} c w - w^{\top} \mu & \\
\mathrm{subject \ to} & 1^{\top} w = 1.&
\end{array}
\end{equation*}
As we can see, this formulation does not depend on the capital but assumes that we invest $100\%$ by constraining $1^{\top} w = 1$. The discretization program in Eq.~\ref{eq:discretization} is solved to obtain a capital-based discrete allocation.

There are two slightly different ways to encode the budget for the discrete optimization cases: the budget can be expressed in terms of assets or in terms of capital. Let us first consider the budget, which is expressed in terms of assets. In other words, this means that the sum $1^{\top}z = \sum_{i}z_i$ for $z = (z_1, \ldots, z_n)$ should be limited. Let $B \in \mathds{Z}_{>0}$ denote the budget in terms of assets, i.e., the number of assets that the investor can buy. After including the budget constraint, the previous unconstrained portfolio optimization in Eq.~\ref{eq:unconstrained_problem} becomes constrained:
\begin{equation*}
\begin{array}{lll}
\mathrm{minimize}  & q_0 z^{\top} c z - z^{\top} \mu & \\
\mathrm{subject \ to} & 1^{\top} z = B, & \\
&z \in \mathds{Z}_{\geq 0}^{n}.&
\end{array}
\end{equation*}
This formulation selects a combination of assets such that their total value is $B$.

The problem with the previous discrete formulation is that it assumes that every asset has the same price, which is a very unrealistic simplification. The budget should be given as an investor's capital, which is a real number that describes the amount of capital the investor can invest in the market. 

Next, we describe how to encode a realistic capital-based budget. Let $C \in \mathds{R}_{>0}$ be the investor's capital. Let $p^{\tau} = (p^{\tau}_i)_{i=1}^{n}$ again denote the vector of closing prices from the last time moment $\tau$, i.e. $p^{\tau}$ are the current prices that the investor uses to buy the stocks. Now the relation between discrete quantities of assets $z$, their recent closing prices $p^\tau$ and the capital $C$ (monetary budget) can be written as follows
\begin{equation}\label{eq:budget_constraint}
    \sum_{i=1}^{n}p^\tau_i  z_i = z^{\top}p^{\tau} = C. 
\end{equation}
Using the capital-based budget, the discrete optimization problem obtains a slightly modified formulation:
\begin{equation*}
\begin{array}{lll}
\mathrm{minimize}  & q_0 z^{\top}c z - z^{\top} \mu & \\
\mathrm{subject \ to} & z^{\top}p^{\tau} = C, & \\
&z \in \mathds{Z}_{\geq 0}^{n}.&
\end{array}
\end{equation*}

In real-life cases, this leads to the problem that there might not be an integer variable solution that satisfies the constraint $z^{\top}p^{\tau} = C$ exactly. We will discuss this problem in more detail later and concretely show how it affects the evaluation of different portfolios.



\subsection{Higher moments}\label{subsec:higher_moments}

Mean-variance portfolio optimization assumes normally distributed returns, which often do not reflect reality. To better estimate risks, researchers have incorporated higher moments, skewness and kurtosis, into the model \cite{Bhandari_Das_2009, Lezmi_Malongo_Roncalli_Sobotka_2018, Harvey_Liechty_Liechty_Muller}. Positive skewness indicates a long right tail and it is desirable as it suggests a higher chance of significant positive returns. In contrast, high kurtosis reflects fat tails and sharp peaks, signaling a greater risk of extreme outcomes, which investors typically avoid. Therefore, we aim to maximize skewness and minimize kurtosis.

Expected returns form a one-dimensional vector, and covariances form a two-dimensional matrix. We define the coskewness of returns as a three-dimensional tensor $S = \left\{ S_{ijk}\right\}_{i,j,k}$ for $i, j, k = 1, \ldots, n$ as
\begin{equation}\label{eq:coskewness}
    S_{ijk} = \frac{\mathds{E}[r_i - \mathds{E}[r_i]]\mathds{E}[r_j - \mathds{E}[r_j]]\mathds{E}[r_k - \mathds{E}[r_k]]}{\sigma_i \sigma_j \sigma_k},
\end{equation}
and co-kurtosis of returns as a four-dimensional tensor $K = \left\{ K_{ijkl}\right\}_{i,j,k,l}$ for $i, j, k, l = 1, \ldots, n$ as follows:
\begin{equation}\label{eq:cokurtosis}
    K_{ijkl} = \frac{\mathds{E}[r_i - \mathds{E}[r_i]]\mathds{E}[r_j - \mathds{E}[r_j]]\mathds{E}[r_k - \mathds{E}[r_k]]\mathds{E}[r_l - \mathds{E}[r_l]]}{\sigma_i \sigma_j \sigma_k \sigma_l}.
\end{equation}
To simplify the notation, we denote 
\begin{align*}
S(z) &:= \sum_{i=1}^{n}\sum_{j=1}^{n}\sum_{k=1}^{n}S_{ijk}z_i z_j z_k \text{ and } \\
K(z) &:= \sum_{i=1}^{n}\sum_{j=1}^{n}\sum_{k=1}^{n}\sum_{l=1}^{n}K_{ijkl}z_i z_j z_k z_l.
\end{align*}
Then, the skewness and kurtosis can be included in the optimization problem:

\begin{equation}\label{eq:final_lp}
\begin{array}{lll}
\mathrm{minimize}  & q_2 K(z) - q_1 S(z) + q_0 z^{\top} c z - z^{\top} \mu & \\
\mathrm{subject \ to} &  z^{\top}p^{\tau} = C & \\
& z \in \mathds{Z}_{\geq 0}^{n}. &
\end{array}
\end{equation}
This approach yields a higher-order problem for more realistic portfolio optimization with integer variables and a budget constraint that reflects the actual capital. Moreover, this formulation is relatively close to the higher-order unconstrained binary optimization format in Eq.~\eqref{eq:hubo_formal}. We must still rewrite the constrained optimization problem as the equivalent unconstrained problem and translate the integer variables into binary variables. We describe this process in the following subsection, connecting this problem and its quantum formulation. Note that the same problem can be expressed in terms of continuous variables as
\begin{equation}\label{eq:final_continuous_program}
\begin{array}{lll}
\mathrm{minimize}  & q_2 K(w) - q_1 S(w) + q_0 w^{\top} c w - w^{\top} \mu & \\
\mathrm{subject \ to} &  1^{\top}w = 1. &
\end{array}
\end{equation}

Finally, we are left to choose the scaling values $q_0$, $q_1$, and $q_2$. Finding suitable values is generally a problem in finance. In this work, we choose
\begin{equation*}
    q_0 = \frac{\mathrm{risk\ aversion}}{2}, \ q_1 = \frac{\mathrm{risk\ aversion}}{6}, q_2 = \frac{\mathrm{risk\ aversion}}{24},
\end{equation*}
which are motivated by the so-called Edgeworth expansion for distributions with higher moments \cite{Balieiro_Filho_Rosenfeld_2004}. We choose $\mathrm{risk\ aversion} = 3$.

\begin{example}\label{ex:example1}
Next, we show an example to demonstrate how the continuous variable algorithm works in practice. The same example data is used throughout the article. We fix two companies: the Walt Disney Company (DIS) and the Travelers Companies, Inc. (TRV). The stock market data covers 10 years, from January 1, 2015, to January 1, 2025. The budget is randomly fixed at $723$. First, we compute the expected returns vector with Eq.~\eqref{eq:expected_returns_geometric}, covariance matrix with Eq.~\eqref{eq:covariance}, coskewness tensor with Eq.~\eqref{eq:coskewness}, and cokurtosis tensors with Eq.~\eqref{eq:cokurtosis}. With this data, we construct the optimization problem defined in program \eqref{eq:final_continuous_program}. Solving the problem employs the Sequential Least Squares Programming, which is the default method in Scipy for these types of problems. The solution to the problem is a weight vector for the assets. In this case, the solution is DIS: $63\%$ and TRV: $37\%$. These continuous weights are discretized with the integer program defined in Eq.~\eqref{eq:discretization}. The final allocation is DIS: $4$ and TRV: $1$ with a left-over budget of $37.4$.
\end{example}
\section{Quantum formulation}

This section reformulates the classical higher-order portfolio optimization problem as a higher-order unconstrained binary optimization problem. Then, we discuss how to map the portfolio optimization problems to spin system energy minimization problems. Finally, we present the QAOA circuit, which is the last step before applying the standard QAOA quantum-classical optimization loop.

\subsection{Encoding budget constraint}

In quantum formulations for optimization problems, an unconstrained cost function is necessary. The constrained portfolio optimization problem can be written as an unconstrained problem by heavily penalizing cases that do not respect the constraint. In the case of portfolio optimization, this means introducing a term that is dependent on the remaining budget. The idea is that if the budget constraint is violated, it increases the value of the cost function, penalizing undesirable solutions. Hence, we include the term $\lambda(z^{\top}p^{\tau} - C)^2$ to the cost function in Eq.~\eqref{eq:final_lp} and obtain the final cost function in terms of integer variables for the higher-order portfolio optimization, which can be written as 
\begin{equation}\label{eq:final_hubo}
    \min_{z \in \mathds{Z}_{\geq 0}^{n}} q_2 K(z) - q_1 S(z) + q_0 z^{\top} c z - \mu^{\top} z + \lambda (z^{\top}p^{\tau} - C)^2, 
\end{equation}
where $\lambda >0$ is a constant, determining the weight of the budget constraint. The constraint for the remaining budget is always non-negative. The penalizing constant $\lambda$ should be sufficiently large so that we favor solutions that invest the entire budget. In this work, we explored $\lambda \in \left\{ 0.001, 0.01, 0.1, 0.9, 1.0, 10, 100, 1000 \right\}$. 

Unfortunately, this formulation does not distinguish between allocations that invest over the budget and allocations that invest under the budget. They are both treated similarly because the term $(z^{\top}p^{\tau} - C)^2$ is symmetric with respect to such solutions. It is possible that investing over the budget might not be possible, and thus, we want to avoid such solutions fully. To address this problem, we refine the formulation further. We adopt the constraint encoding idea from \cite{10.3389/fphy.2014.00005}. Instead of considering a budget that is a real number, we approximate it by the closest integer. Let $M = \lfloor \log_2(C)\rfloor$, where $C$ is the the budget capital. Then, we can rewrite the budget as
\begin{equation}\label{eq:budget_constraint_with_slacks}
\sum_{n = 0}^{M - 1}2^{n}y_n + (C + 1 - 2^M)y_M,
\end{equation}
where $y_1, \ldots, y_M$ are binary variables. Then, the penalty term that theoretically prevents us from investing more than we have in the budget is
\begin{displaymath}
\lambda (z^{\top}p^{\tau} - \sum_{n = 0}^{M - 1}2^{n}y_n - (C + 1 - 2^M)y_M)^2.
\end{displaymath}
The idea is to activate and deactivate the binary variables $y_1, \ldots, y_M$ depending on how much budget is used. However, the optimal point will never include allocations where we would invest more than the budget. Although this encoding adds only a logarithmic overhead to the encoding, it easily creates portfolio optimization problem instances with a relatively large number of qubits. Hence, we continue with the first encoding in Eq.~\eqref{eq:final_hubo} so that both cases are penalized equally. We will discuss the effects of this choice later in the discussion section.

\subsection{Variable encoding in quantum formulation}\label{subsec:variable_encoding}

Since we are developing the portfolio optimization on quantum computers, we must use binary variables. To use integer variables that are encoded with binary variables, we employ a similar method as in Eq.~\eqref{eq:budget_constraint_with_slacks}, which is the standard variable rewriting method from \cite{10.3389/fphy.2014.00005}. The same encoding was also used in \cite{Buonaiuto_Gargiulo_De_Pietro_Esposito_Pota_2023}. The method translates integer variables into binary variables with a logarithmic overhead. Let $z$ be an integer variable that obtains values at range $0, \ldots, N$. First, choose $M = \lfloor{\log(N)} \rfloor$ with base $2$ which implies $2^{M} \leq N \leq 2^{M + 1}$. Then, instead of using the naive encoding with $y_1, \ldots, y_M$ binary variables, we can employ the logarithmic encoding as
\begin{equation}\label{eq:int_to_bin_substitution}
z = \sum_{n = 1}^{N} ny_n \to (N + 1 - 2^M)y_M + \sum_{n = 0}^{M - 1}2^{n}y_n.
\end{equation}

We must know the range $0, \ldots, N$ for the integer variables to rewrite the integer variables as binary variables. In portfolio optimization problems, this range can be computed from the latest prices and the budget. For each integer variable $z_i$ that is assigned to an asset $i$, we have the latest price $p^{\tau}_i$. Then, the maximum amount of assets we can buy is $\lfloor p^{\tau}_i/B \rfloor$, i.e., investing the whole budget to a single company, providing us with the range. We obtain the final higher-order unconstrained binary optimization problem when this variable transformation is applied to the problem formulation in Eq.~\eqref{eq:final_hubo}.

\subsection{From optimization to spin system energy minimization}\label{subsec:spin_system}

Following the binary formulation, the previously described portfolio optimization cost function in Eq.~\eqref{eq:final_hubo} is mapped to the Hamiltonian minimum eigenvalue problem. The goal is to find a configuration of spin variables that minimizes the energy of the whole system. In the portfolio optimization problem, the Hamiltonian can be written using Pauli-Z operators that have eigenvalues $\pm1$. In binary optimization problems, using the mapping $x_i = \frac{1}{2} (1-s_i)$, we map the binary variables $x_i \in \{ 0, 1\}$ to the spin variables $s_i \in \left\{-1, 1\right\}$. Therefore, the portfolio optimization problem can be written as finding the minimum eigenvalue of the Hamiltonian corresponding to the lowest cost function value.

\subsection{QAOA circuit}\label{subsec:cost_hamiltonian}

\begin{figure}[t]
    \centering
    \input{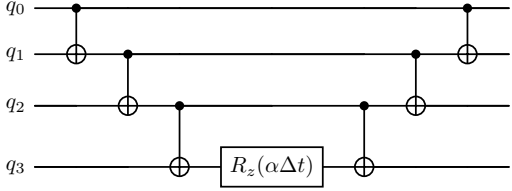}
    \caption{Circuit block encoding higher-order term $\alpha Z_0 \otimes Z_1 \otimes Z_2 \otimes Z_3$}
    \label{fig:qaoa_hubo}
\end{figure}

In this subsection, we describe how to construct the parametrized QAOA circuit from the Hamiltonian that describes the higher-order unconstrained binary optimization problem in Eq.~\eqref{eq:final_hubo}. Let $I$ be the indexing set, i.e., in this case, the set containing the qubits. Currently, the Hamiltonian has the following form
\begin{equation}
    H = \sum_{S \subset I}\alpha_{S}\prod_{i \in S}Z_i.
\end{equation}
This is necessarily the same formulation that we expressed for abstract HUBO problems in Eq.~\eqref{eq:hubo_formal}. In this higher-order portfolio optimization case, the size of the sets satisfies $|S| \leq 4$, which follows from the dimension of the cokurtosis tensor. The other typical case is the QUBO problems where $|S| \leq 2$. As described in \cite{Nielsen_Chuang_2010}, the quantum circuit implementing $e^{-iH\Delta t}$ is relatively easy to construct, and the idea generalizes the method for QUBO problems. For example, consider that the Hamiltonian $H$ contains a term $\alpha Z_0 \otimes Z_1 \otimes Z_2 \otimes Z_3$. This becomes the circuit block that is visualized in Fig.~\ref{fig:qaoa_hubo}.

Additionally, the QAOA circuit consists of the mixer layer, and we employ the standard $x$-mixer. Finally, the expectation value of the cost Hamiltonian is measured. Modern quantum computing frameworks, such as Qiskit and Pennylane, perform the QAOA circuit construction and parametrization automatically, even for higher-order problems.

\begin{example}
We continue Example \ref{ex:example1} by creating the corresponding higher-order binary optimization problem for the portfolio optimization problem, which includes Disney and Travelers Company. The closing prices for the companies are DIS: $111$ and TRV: $240$. The budget is $723$. If the entire capital is allocated to DIS, it is possible to buy at most $\lfloor 723/111\rfloor = 6$ assets of Disney and, similarly, $\lfloor 723/240\rfloor = 3$ assets of Travelers. At least $3$ qubits are needed to encode the integers in the interval $[0,6]$ for Disney and $2$ qubits for integers in interval $[0,3]$ for Travelers. Thus, the problem is encoded with $5$ qubits.

Again, the expected returns, covariances, coskewness, and cokurtosis are computed. Based on this data, the corresponding unconstrained integer variable optimization problem described in Eq.~\eqref{eq:final_hubo} is constructed. Then, the integer variables are translated into binary variables by employing the substation in Eq.~\eqref{eq:int_to_bin_substitution}. For example, for Disney, the following substitution is performed at every position in the formulation
\begin{equation*}
    z \to x_0 + 2x_1 + 4x_2,
\end{equation*}
where $z$ is the integer variable in interval $[0,6]$ and $x_0$, $x_1$and $x_2$ are binary variables. Three binary variables, i.e., three qubits, encode a single integer variable $z$. Again, these binary variables $x_0$, $x_1$ and $x_2$ representing the integer $z$ are mapped into the spin glass ground state energy problem by using $x_i \leftrightarrow (1 - Z_i)/2$ as explained earlier. This is the final form for the Hamiltonian that encodes the higher-order portfolio optimization problem.

As in the standard QAOA, the higher-order problem becomes a parametrized quantum circuit. For this example, the circuit is partially visualized in Fig.~\ref{fig:qaoa_circuit_example}. The figure includes an example of all terms with varying degrees from the first to fourth degree. The corresponding Pauli matrices from the final Hamiltonian are visualized on top of each term. The values in $R_z$-gates are obtained from the stock data and the initial parameter values for the algorithm.

\begin{figure*}[t]
    \centering
    \input{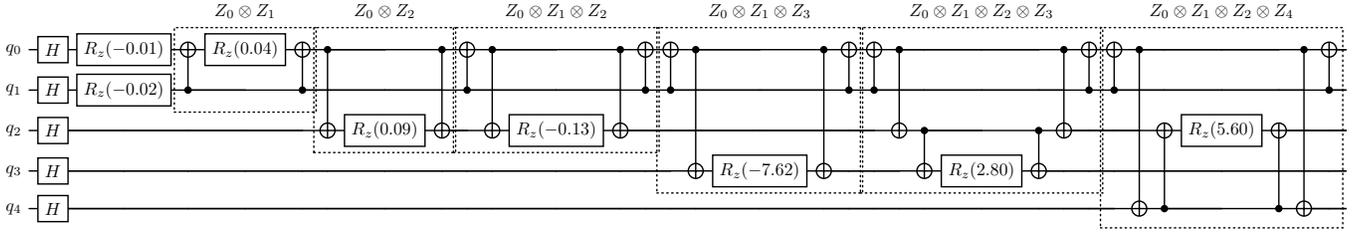}
    \caption{Part of the QAOA circuit}
    \label{fig:qaoa_circuit_example}
\end{figure*}

For this example, the solution for the corresponding higher-order binary optimization problem is DIS: $0$ and TRV: $3$, and the left-over budget is $2.9$. The problem is solved by finding the eigenvalues and eigenvectors of its Hamiltonian. In this particular case, QAOA could also find this optimal point.

\end{example}

\section{Experimental setting}
This section describes the experimental setup that demonstrates the utility of the proposed algorithm compared to the classical baselines and mean-variance portfolio optimization. We describe the stock market data used to create the optimization problems, followed by the classical baseline implementations and the employed optimizers.

\subsubsection{Data sets}

We used the Dow Jones Industrial Average (DJIA), a stock market index that tracks 30 large, publicly traded U.S. companies. These companies represent a diverse range of industries, including technology, finance, healthcare, and consumer goods. The tickers (i.e., the shorthand stock market names) for the companies were downloaded from the Python library \emph{pytickersymbols} \cite{pytickersymbols}, and the stock market data was downloaded from \emph{yfinance} \cite{yfinance}. We downloaded data from January 1, 2015, to January 1, 2025, spanning ten years. 


Next, we describe how we have constructed the portfolio optimization problems. We randomly sampled between $2$ and $10$ companies from the $30$ companies in the index. Then, we randomly assigned a budget, which we limited to $6000$. If the budget exceeded $6000$, it was unlikely that the corresponding higher-order QAOA circuit for the portfolio optimization problem would have been simulable with the available resources. We calculated the number of qubits required for this portfolio optimization case. If the number of qubits was over $15$, we excluded the optimization case from the dataset. Otherwise, it was included until every case from $6$ to $15$ qubits had ten random problems. This way, we obtained $100$ random portfolio optimization problems that were ensured to fit a 15-qubit simulator. 

\begin{figure}[t]
    \centering
    \includegraphics[width=0.7\linewidth]{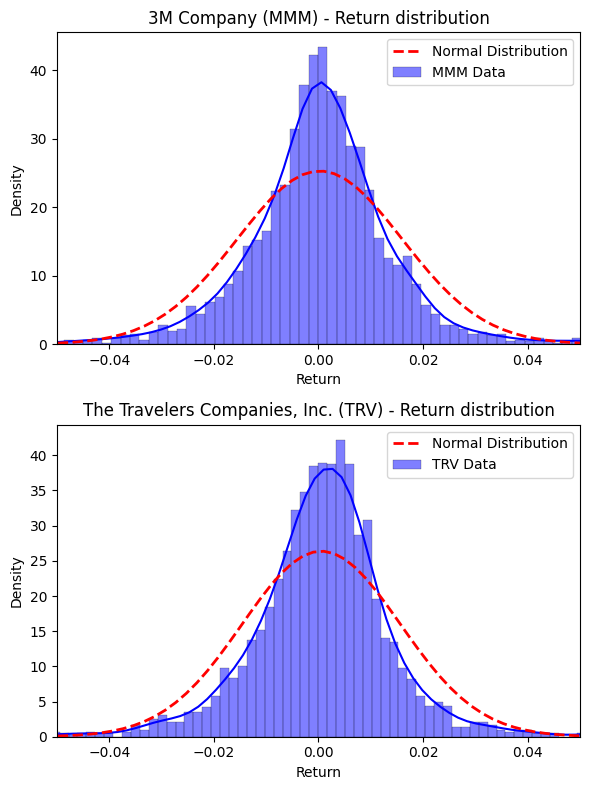}
    \caption{Return distributions for two companies with one of the highest skewness and kurtosis terms.
    The actual distribution of companies' return values is presented in purple, and the normal distribution is denoted as a red dotted line. From the figures, it is clear that the assumption of a normal distribution of returns is often inaccurate; therefore, higher-order portfolio optimization may be a better approach to modeling the problem. }
    \label{fig:3m_trv}
\end{figure}

Next, we argue that the stock market data included in the experimental dataset is not normally distributed, meaning it exhibits skewness and kurtosis. This motivates us to include the higher moments in the optimization problem. Fig.~\ref{fig:3m_trv} demonstrates returns distributions for two companies, 3M and Travelers. Compared to the normal distribution, both distributions show clear positive kurtosis, i.e., the high peak. The distribution for Travelers also exhibits negative skewness, as the distribution has shifted to the right. Later in the discussion section, we will also explore how including higher-order moments produces more diverse portfolios, which can be considered a positive argument for including those moments.

\subsubsection{Classical algorithms and optimizers}

We employ classical optimizers in three different ways: classical optimizers solve the classical continuous variable baseline, another classical optimizer is used in the classical subroutine of QAOA, and a standard eigenvalue solver computes the exact spectrum for the HUBO problems when we have fewer than 14 qubits. When the problem consists of 14 or 15 qubits, we used the sparse eigenvalue solver, which does not return the whole spectrum but only approximates the smallest eigenvalues.

First, the classical continuous variable baseline utilizes SLSQP (Sequential Least Squares Programming), which is the default optimizer for SciPy if the optimization problem has constraints and bounds. Then, an integer program turns the continuous variables into discrete allocations. The integer program is solved with CVXPY \cite{diamond2016cvxpy,agrawal2018rewriting}.

\begin{figure}[t]
    \centering
    \includegraphics[width=0.99\columnwidth]{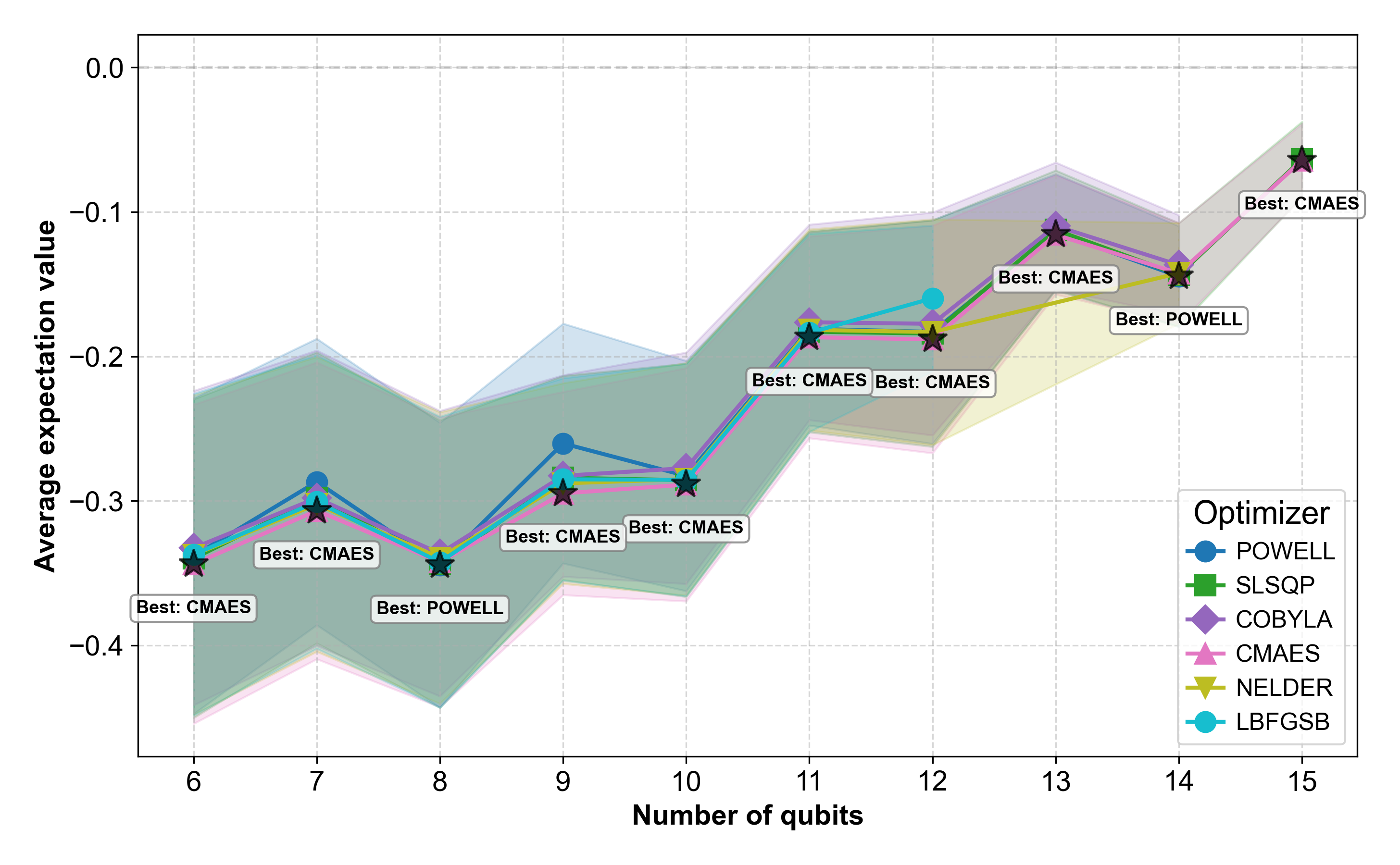}
    \caption{Comparison between the different classical optimizers in optimizing the QAOA circuits solving the HUBO problem. Shaded areas represent 95\% confidence intervals around the mean. The results are only presented for optimizers that can complete at least 8 of 10 problems for each qubit count.}
    \label{fig:average_expectation_value_by_qubit_count}
\end{figure}

Second, we benchmarked six optimizers (Powell, SLSQP, COBYLA, CMA-ES, Nelder-Mead, and L-BFGS-B) in the QAOA's subroutine. A brief comparison of the optimizers for solving higher-order portfolio optimization with QAOA is presented in Figure~\ref {fig:average_expectation_value_by_qubit_count}. The lowest average objective function values can often be achieved with the Covariance Matrix Adaptation Evolution Strategy (CMA-ES) \cite{Hansen_2023,hansen2019pycma,hansen2001ecj}, but this comparison lacks, for instance, noise and hyperparameter search. Nevertheless, we chose to investigate the performance of the QAOA algorithm with a CMA-ES optimizer. The optimizer is a stochastic gradient-free numerical optimization algorithm for complex (non-convex, ill-conditioned, multi-modal, rugged, and noisy) optimization problems in continuous search spaces. We have not previously seen it employed as an optimizer for QAOA. We chose $\sigma = 0.1$ for the CMA-ES algorithm and limited the number of iterations, but the other hyperparameters were left to their default values.

Finally, we briefly describe the classical baselines. The classical baseline is to use the continuous variable higher-order program in Eq.~\ref{eq:final_continuous_program}. As noted earlier, this formulation is solved so that it does not invest over the budget, i.e., the budget constraint is strictly satisfied. On the other hand, the quantum formulation does not strictly satisfy the budget constraint because the constraint is part of the objective function. Thus, we have included the unconstrained continuous variable formulation
\begin{equation}\label{eq:final_hubo_continuous}
    \min_{w \in [0,1]^n} q_2 K(w) - q_1 S(w) + q_0 w^{\top} \Sigma w - \mu^{\top} w + (1^{\top}w - 1)^2, 
\end{equation}
which is closer to the HUBO formulation. Both continuous variable solutions are discretized with the program in Eq.~\eqref{eq:discretization}.

\subsubsection{Evaluation metric}

The integer optimization problem was described in Eq.~\eqref{eq:final_lp}. It consists of two parts: the objective function
\begin{equation}\label{eq:objective}
f(z) = -q_2 K(z) + q_1 S(z) - q_0 z^{\top} c z + \mu^{\top} z,
\end{equation}
and the budget constraint. The objective $f(z)$ provides a value that can be used to compare different solutions. On the other hand, in real-life portfolio optimization, we rarely find a portfolio allocation with the same value as the budget. Thus, the realistic final result always breaks the budget constraint, and this difference in budget should penalize the solution. Hence, the final evaluation for a solution $z$ should be based on the value obtained from the objective function $f(z)$ and the difference in budget denoted by $d_i$.

Each method produces a discrete portfolio allocation $z$. The values from the cost function $f(z)$ without the budget constraint vary greatly between different portfolio optimization cases. Thus, we employ min-max normalization, which is commonly used in machine learning. This normalizes the value of $f(z)$ to the interval $[0,1]$ and is defined
\begin{equation*}
    f_{\mathrm{norm}}(z) = \frac{f(z) - \min\left\{ f(z_i) \mid i \in I \right\}}{\max\left\{ f(z_i) \mid i \in I \right\} - \min\left\{ f(z_i) \mid i \in I \right\}},
\end{equation*}
where $I$ is the set containing the indices for different methods, i.e., classical methods, QAOA, and exact eigensolver, and $z_i$ are the optimized allocations for each method $i \in I$.
\section{Results}

\begin{figure*}[t]
    \centering
    \begin{subfigure}{0.33\textwidth}
        \centering
        \includegraphics[width=\linewidth]{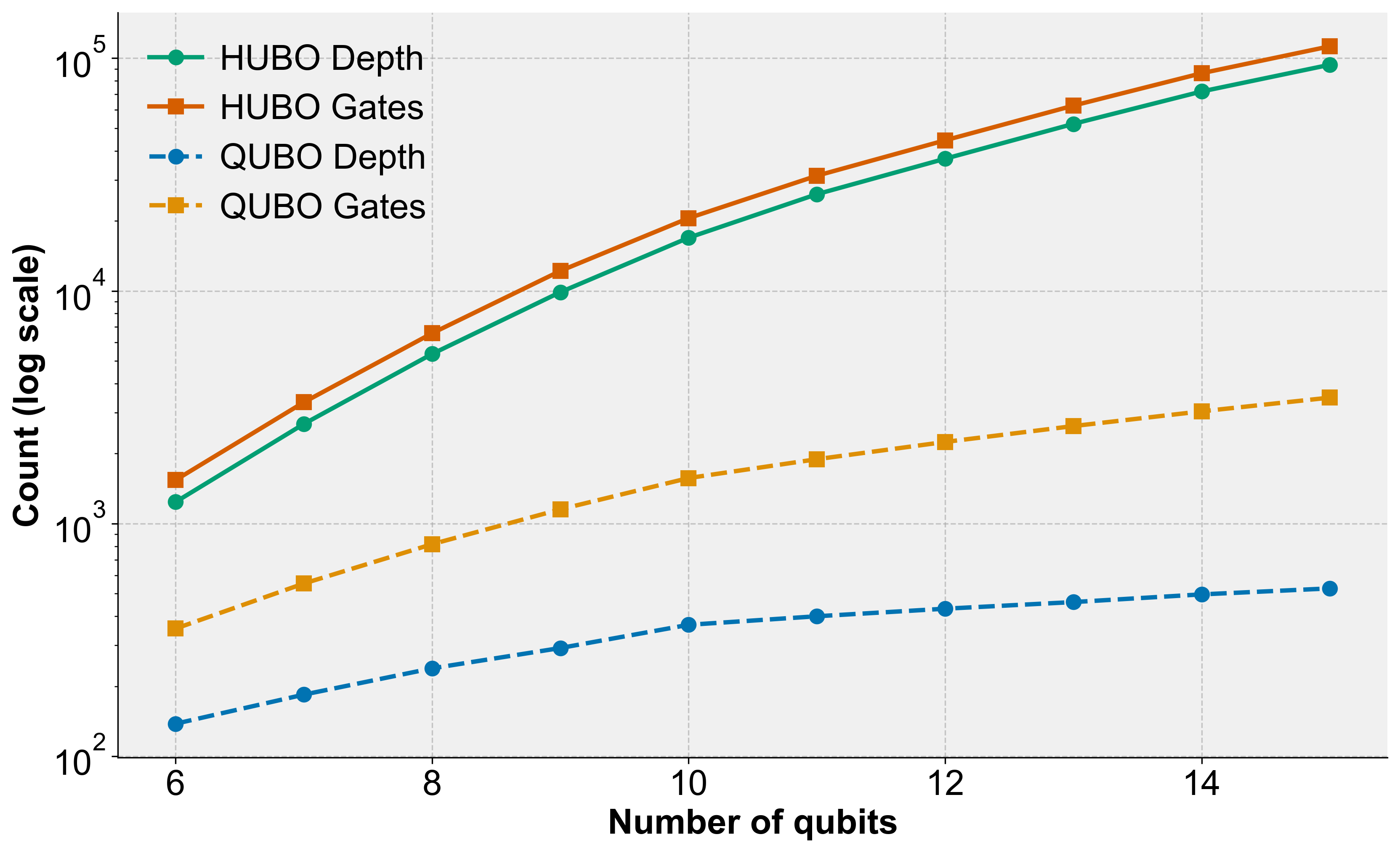}
        \caption{Gate count and depth comparison for QUBO and HUBO circuits in QAOA.}
        \label{fig:circuit_metrics}
    \end{subfigure}
    \begin{subfigure}{0.32\textwidth}
        \centering
        \includegraphics[width=\linewidth]{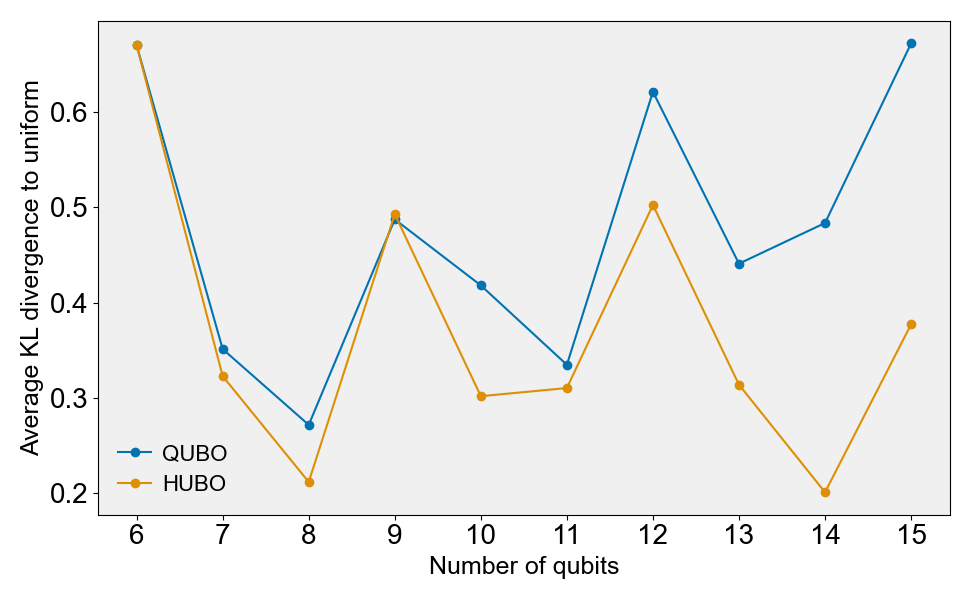}
        \caption{Average KL divergence of QUBO and HUBO allocations compared to uniform.}
        \label{fig:average_kl_divergence}
    \end{subfigure}
    \begin{subfigure}{0.33\textwidth}
        \includegraphics[width=\linewidth]{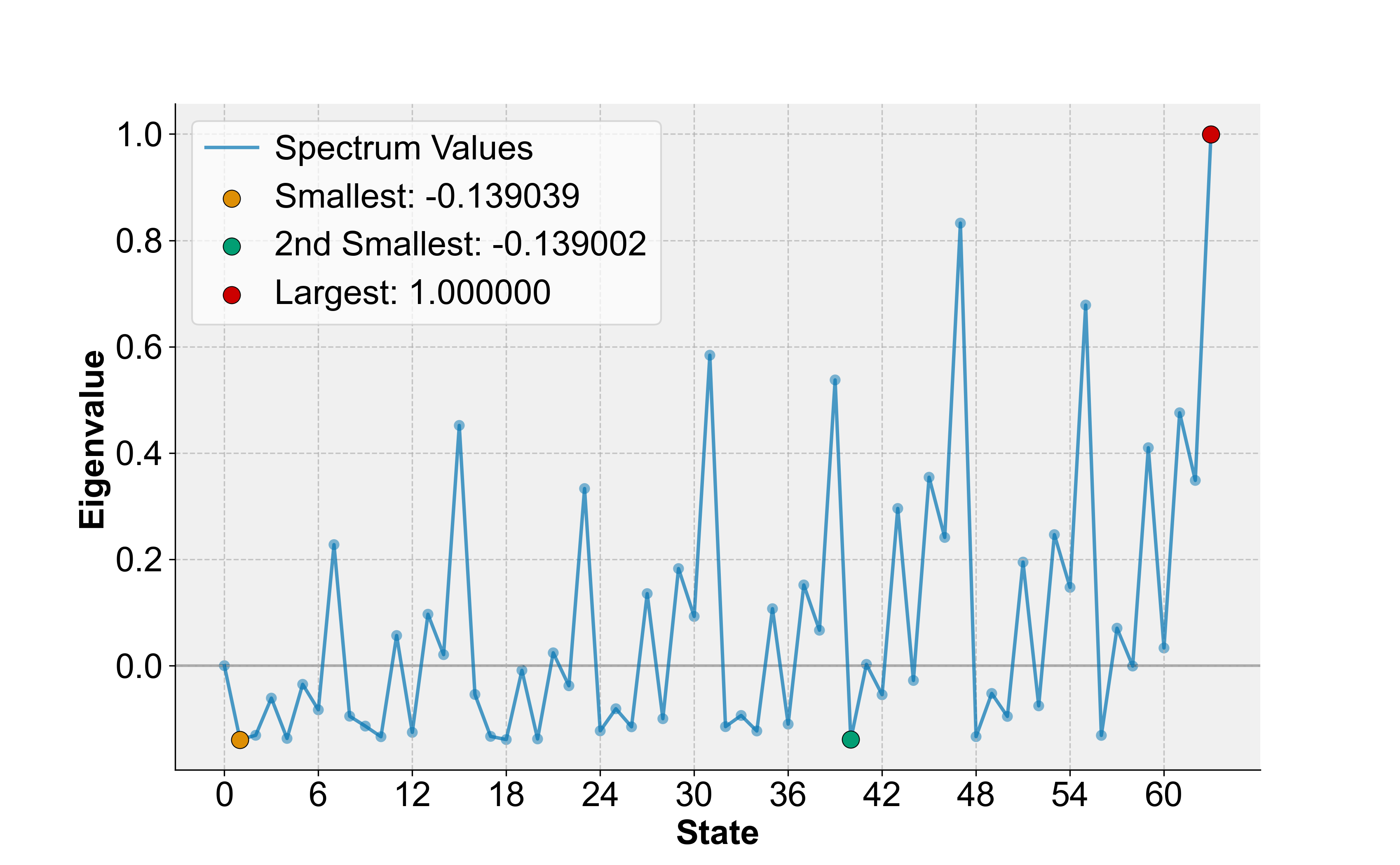}
    \caption{Eigenvalues for a higher-order portfolio optimization problem with $6$ qubits.}
    \label{fig:spectrum_plot_6}
    \end{subfigure}
    \caption{Comparison of circuit metrics and allocation divergences for QUBO and HUBO settings.}
    \label{fig:combined_figure}
\end{figure*}

In this section, we discuss our findings and visualize the problem characteristics that explain QAOA's performance in comparison with the exact solution and classical methods in this higher-order portfolio optimization problem. Due to space limitations, we have included the results from the smallest qubit counts, ranging from 6 qubits to 8 qubits, and the largest qubit counts, ranging from 12 to 15 qubits. The complete set of detailed results can be viewed on GitHub~\cite{github_repo}. 

The selection of $\lambda$-values, which balances between the objective function in Eq.\ref{eq:objective} and the budget constraint, affects the performance of QAOA. We performed the same optimization setup with varying  $\lambda$ values (0.001, 0.01, 0.1, 1.0, 0.9, 10, 100, and 1000). In the results, we present the QAOA performance with the most suitable $\lambda$. Due to space limitations, we do not visualize the optimal $\lambda$ for each optimization problem. However, generally, values higher than $1$ performed the best.

We also briefly discuss how much the QAOA algorithm improves the probability of measuring the state that we consider as a solution compared to the randomly sampled state. For each qubit we compute a so-called QAOA enhancement factor which is the observed highest probability after applying QAOA divided by the uniformly random probability of $1 / 2^{n}$, where $n$ is the number of qubits. The average enhancement across all cases is $14.43$, which means that on average it is $14.43$x more probable to measure the minimizing state compared to a random state. The worst enhancement factor we observed was $3.79$, and the largest was $108.94$.

\subsection{Results compared to the classical baseline}

\begin{figure*}
    \begin{subfigure}[b]{0.49\textwidth}
        \includegraphics[width=\textwidth]{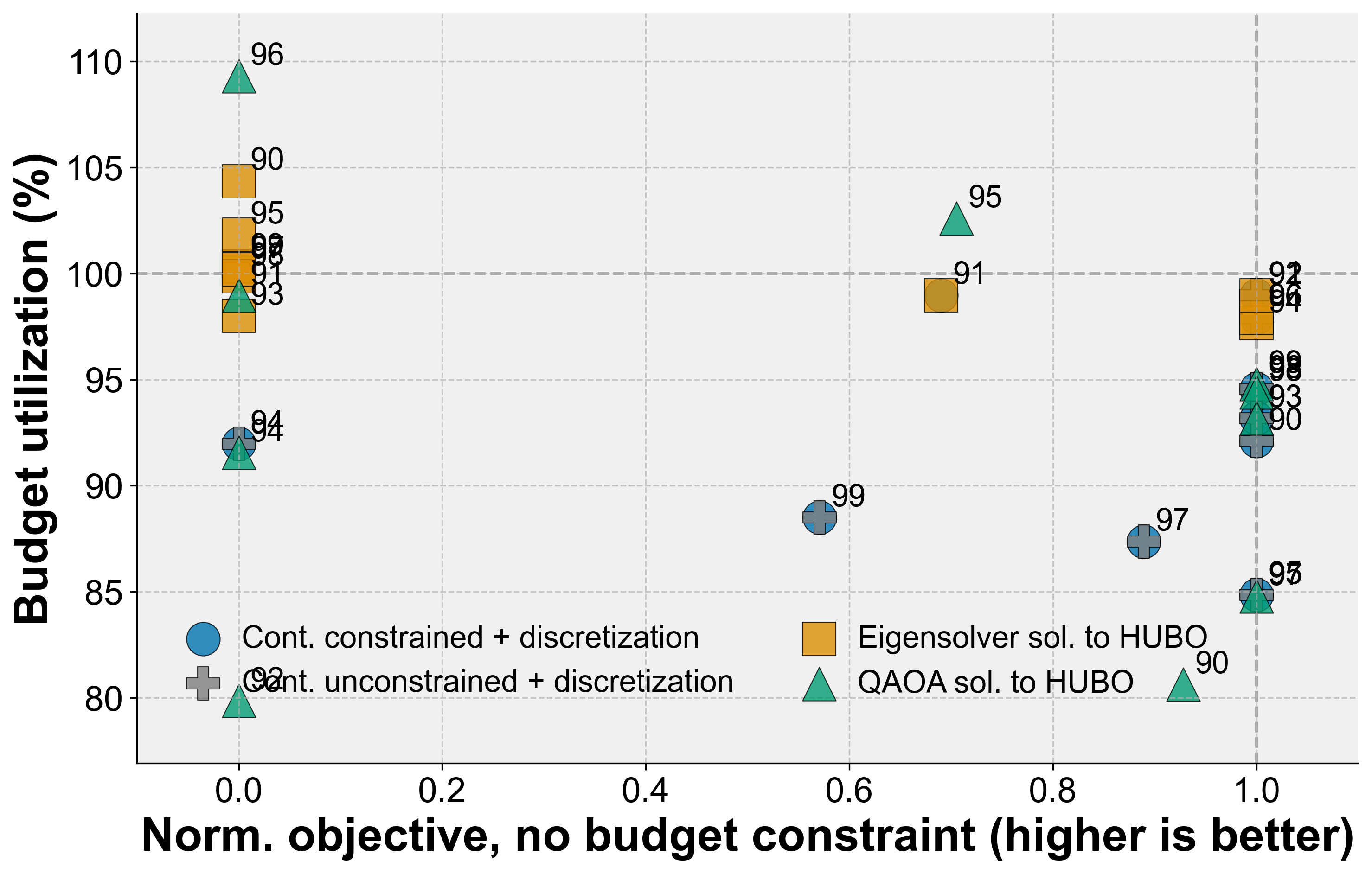}
        \caption{Portfolio optimization cases with $6$ qubits}
        \label{fig:6_qubits}
    \end{subfigure}
    \begin{subfigure}[b]{0.49\textwidth}
        \includegraphics[width=\textwidth]{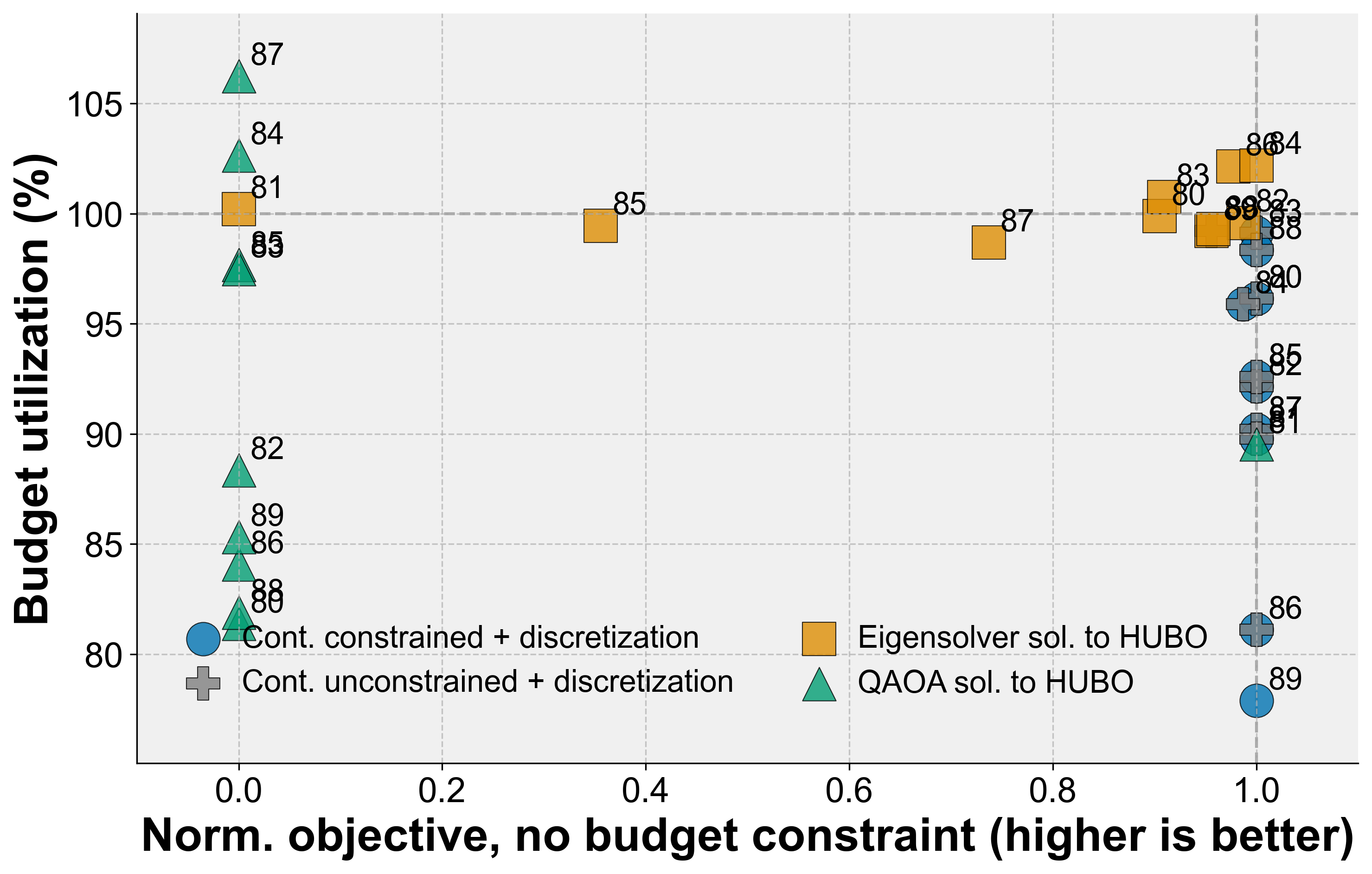}
        \caption{Portfolio optimization cases with $7$ qubits}
        \label{fig:7_qubits}
    \end{subfigure}
    \hfill
        \begin{subfigure}[b]{0.49\textwidth}
        \includegraphics[width=\textwidth]{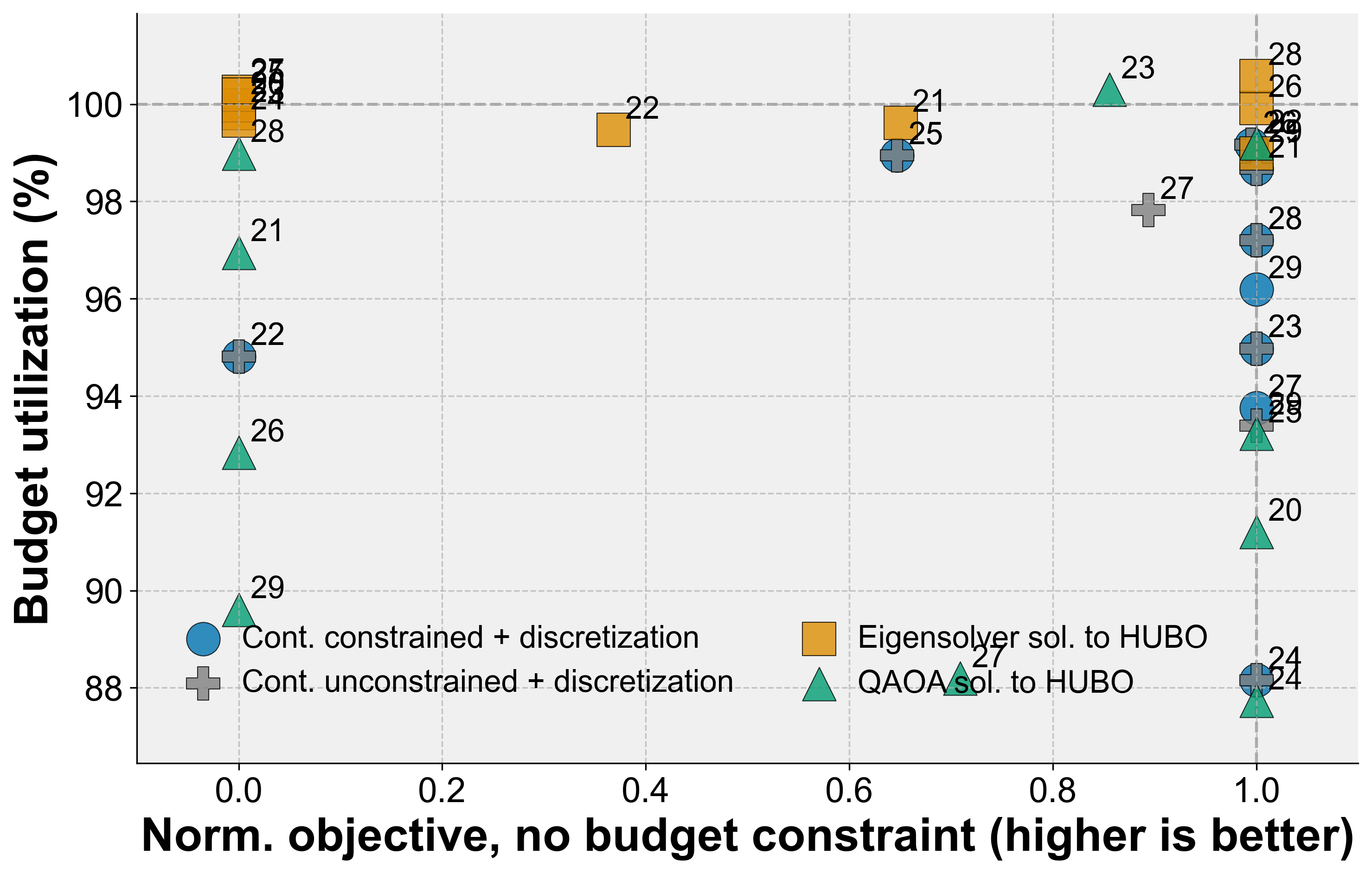}
        \caption{Portfolio optimization cases with $8$ qubits}
        \label{fig:8_qubits}
    \end{subfigure}
    \begin{subfigure}[b]{0.49\textwidth}
        \includegraphics[width=\textwidth]{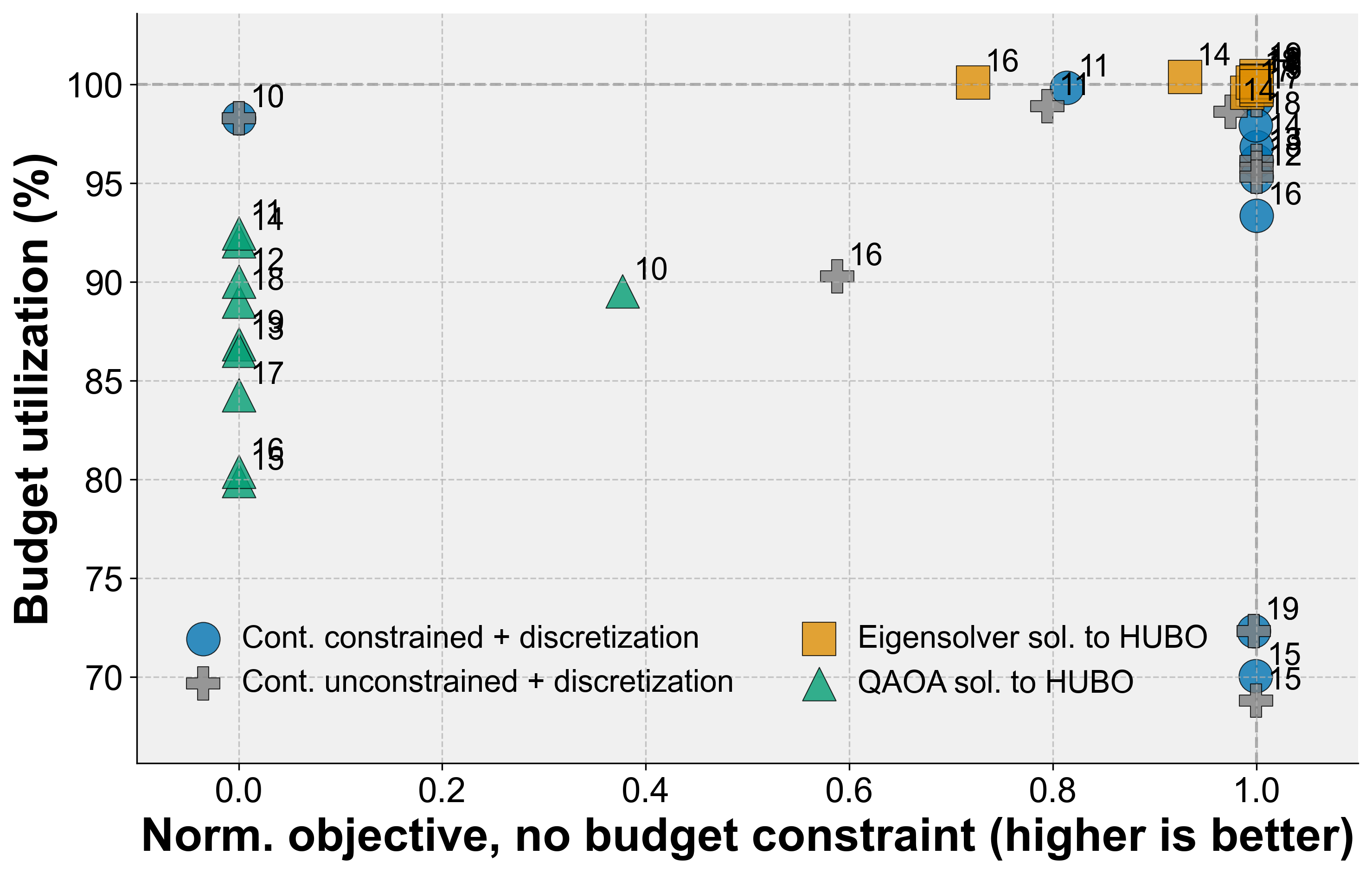}
        \caption{Portfolio optimization cases with $13$ qubits}
        \label{fig:13_qubits}
    \end{subfigure}
    \hfill
        \begin{subfigure}[b]{0.49\textwidth}
        \includegraphics[width=\textwidth]{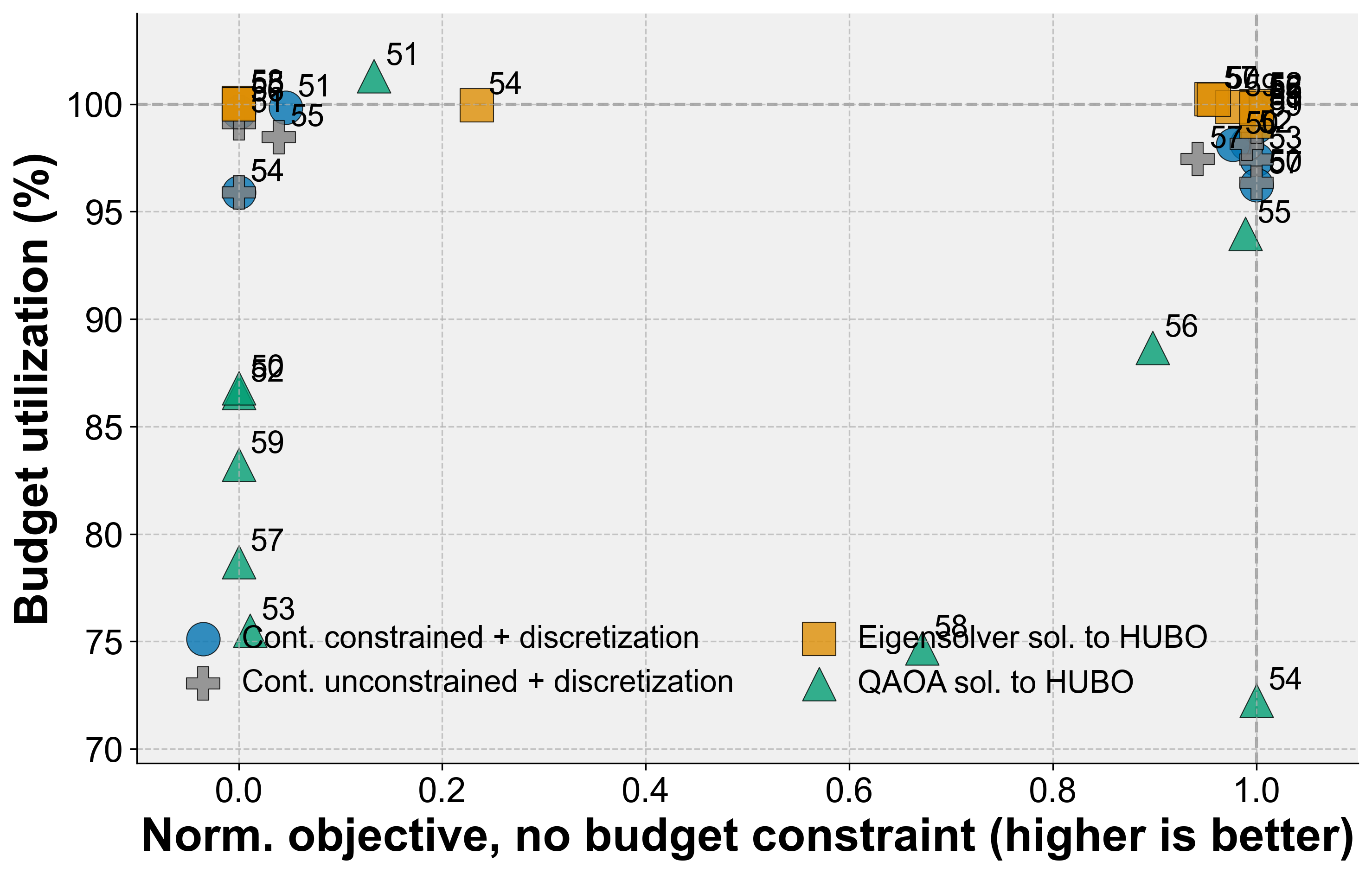}
        \caption{Portfolio optimization cases with $14$ qubits}
        \label{fig:14_qubits}
    \end{subfigure}
    \begin{subfigure}[b]{0.49\textwidth}
        \includegraphics[width=\textwidth]{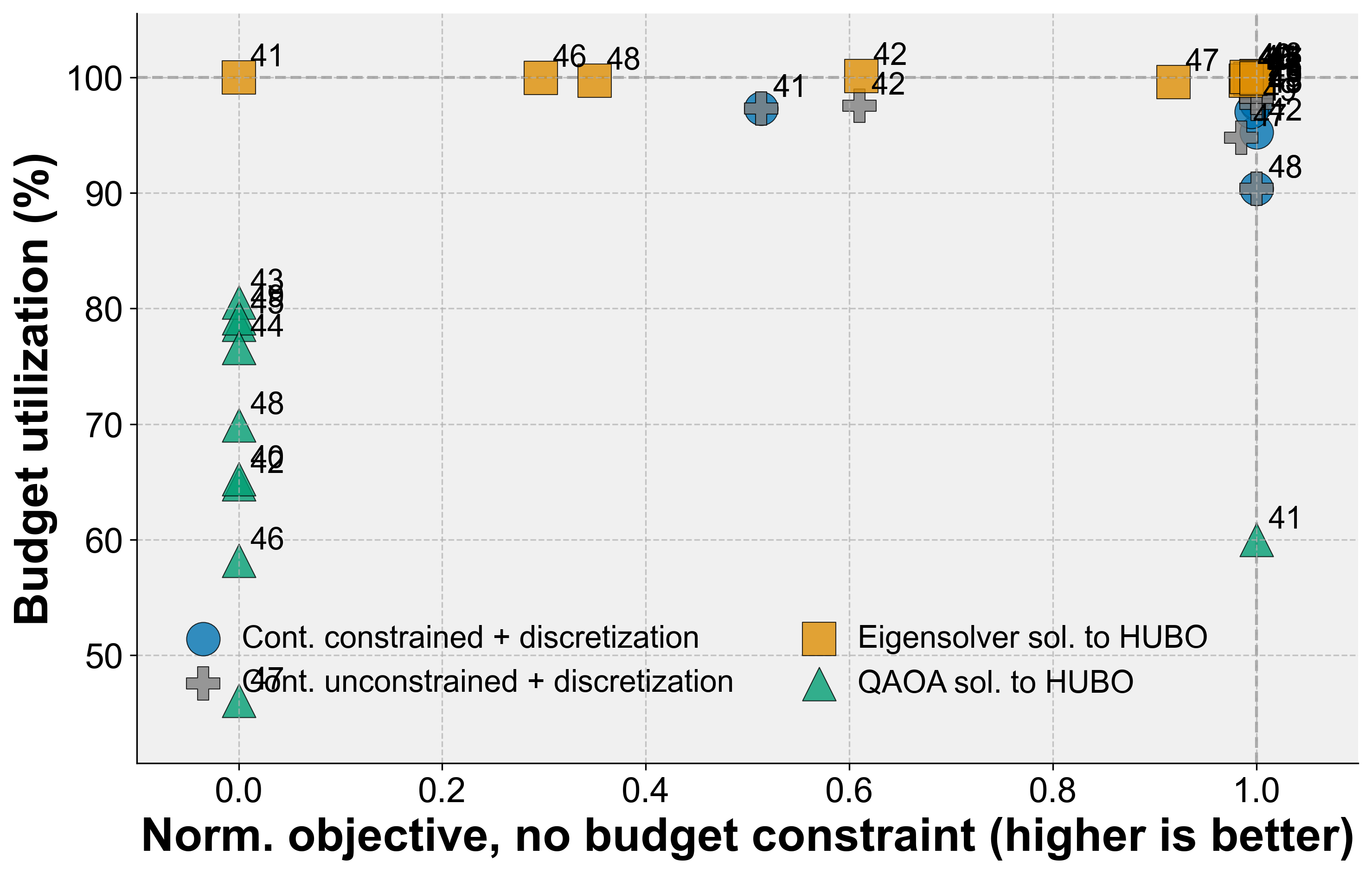}
        \caption{Portfolio optimization cases with $15$ qubits}
        \label{fig:15_qubits}
    \end{subfigure}
    \caption{Portfolio optimization cases with various qubit counts}
    \label{fig:qubits_12_to_15}
\end{figure*}

The results of solving selected portfolio optimization problems are presented in Fig. \ref{fig:qubits_12_to_15}, where budget utilization is shown on the y-axis, and the x-axis shows the min-max normalized value of the cost function $f(z)$ without the budget constraint. The best solutions are in the upper right corner and along the 100\% line. One of the findings is that our higher-order formulation for the portfolio optimization problem is often theoretically capable of encoding portfolio allocations whose quality outperforms that of solutions from classical baselines, as seen in Fig.~\ref{fig:qubits_12_to_15} and in Table~\ref{tab:algorithm_comparison}. This result motivates further study of methods to solve HUBO problems on quantum hardware. The solutions to the HUBO problems have aimed to invest the whole budget, and this goal has sometimes come at the cost of smaller expected returns. Unfortunately, solving higher-order portfolio optimization problems with QAOA proved challenging, and only a small subset of problems matched the quality of the solutions obtained with exact methods or classical algorithms. This suggests that QAOA may require modifications to the optimization pipeline before it can effectively address more complex HUBO optimization problems.

\begin{table}[htbp]
\caption{Results of comparing the optimization approaches}
\label{tab:algorithm_comparison}
\centering
\resizebox{\columnwidth}{!}{
\begin{tabular}{l c c c c}
\toprule
\textbf{Metric} & \textbf{Constrained} & \textbf{Unconstrained} & \textbf{QAOA} & \textbf{HUBO} \\
& \textbf{classical} & \textbf{classical} & & \textbf{exact} \\
\midrule
$95\% \leq$ budget util. $\leq 105\%$ & 74 & 70 & 19 & \textbf{100} \\

min-max objective $\geq 0.95$ & \textbf{70} & 68 & 23 & 53 \\

Intersection of both & 50 & 46 & 3 & \textbf{53} \\
\bottomrule
\end{tabular}
}

\end{table}

We suspect that the QAOA's performance may be limited by the complex optimization landscape of portfolio problems, which often have many local minima. The current results demonstrate that the optimization becomes more challenging as the number of qubits increases. Fig.~\ref{fig:spectrum_plot_6} shows the eigenvalue spectrum for a 6-qubit instance, with several values near the optimum, suggesting a rugged landscape that makes these problems challenging for optimization. It will be a future research topic to improve the methods so that QAOA can tackle HUBO problems more efficiently.


\subsection{Comparison to mean-variance portfolio optimization}

Our implementation \cite{github_repo} enables users to solve the same portfolio optimization problems using either the mean-variance formulation, which is a QUBO problem, or the higher-order formulation, which is an HUBO problem. This subsection compares the HUBO problem to the corresponding QUBO problem. This clarifies the effect of including the higher-order terms and the complexity of the HUBO problem. 

The first comparison is between the required gate counts and depths for each problem, which is presented in Fig.~\ref{fig:circuit_metrics}. We did not perform optimization routines for the circuits, and the gates used are the standard for QAOA: Hadamard, CNOT, $R_z$, and $R_x$ rotations. Both problems demonstrate scalability, which remains a challenge for current real quantum computing hardware.

Next, we compare the spectra of each problem by computing the exact spectra for HUBO and QUBO problems in the cases between 6 and 13 qubits and calculating the variance of the differences. This showed that for small qubit counts, the difference in the spectra between HUBO and QUBO problems is very small at level $10^{-9} - 10^{-11}$. This reveals that with the current weights for coskewness and cokurtosis, their effect on the values of the spectrum is small. The difference in the spectra becomes more dominant with increasing qubit counts.

Although the spectra for these optimization problems resemble each other, the allocations start differing substantially when the number of qubits and complexity increase. We visualize the difference in the allocations by computing how much they differ as distributions and how far they are from the allocation, which would invest an equal amount in every asset. The results are in Fig.~\ref{fig:average_kl_divergence}. KL divergence, i.e., relative entropy, is a standard metric for comparing two distributions. The lower KL divergence value in the figure indicates that the distribution is closer to the uniform distribution. The closer the two plots are, the more similar allocations they provide on average. \looseness=-1

The results in Fig.~\ref{fig:average_kl_divergence} show that when the number of qubits increases, the difference between QUBO and HUBO allocations becomes more significant. We can also read that the HUBO solutions stay closer to the uniform allocation, which means that this formulation produces more diverse portfolios compared to the mean-variance problem. We consider portfolio diversification to be a positive feature.
\section{Conclusion}






Despite being a complex problem with a strong connection to real-life data, portfolio optimization has a central role in quantum optimization. In this article, we introduced a higher-order portfolio optimization problem and showed how it can be formulated for quantum computers and solved with QAOA. We demonstrated its performance against the classical baselines and showed that the HUBO formulation can encode better portfolios than the classical solutions. This work also concretely demonstrated differences between mean-variance and higher-order portfolio optimization and briefly compared the most common classical optimizers. While the resource requirements for solving HUBO problems with QAOA are likely too demanding for the current hardware, HUBO problems show characteristics that make them a promising class of problems to be solved on quantum computers, especially because the classical solvers do not natively support them. 

Future research will include improving QAOA, optimization strategies, and classical subroutines to work more efficiently for HUBO problems. We are especially interested in exploring tensor networks and Bayesian optimizers.



\section{Acknowledgement}
This work is funded by Business Finland (grant number 169/31/2024), Research Council of Finland (grant number 362729) and the Quantum Doctoral Programme.

\FloatBarrier
\balance
\bibliographystyle{IEEEtran}
\bibliography{ref.bib}

\end{document}